# The transverse wakefield calculated by the double circuit model

Yang Xiong-Bin(杨雄斌)[1)] Hou Mi(侯汨)[2)]
1 University of Chinese Academy of Sciences, Beijing 100049, China
2 Institute of High Energy Physics, Chinese Academy of Sciences, Beijing 100049, China

Abstract: X-band accelerators for multi-bunches are a new way to produce high luminosity and energy efficiency bunches. Smaller the size and more bunches, more severe is the wakefield in the X-band accelerators, unless some means of strongly suppressing the transverse wakefield is adopted in the design of the accelerating structure. Here, the derivation of wakefield function of the double circuit model and its application to the accelerator structure designed have been demonstrated.

1. Introduction

For multi-bunch accelerating structures, the long range transverse wakefield is one of the most important problem which needs to be addressed, as it has bad effect to the transverse emittance. To date, the transverse wakefield can be calculated by four models: uncoupled model, single-band equivalent-circuit model, double-band equivalent-circuit model and spectral function method. The calculation of wakefield by the double circuit model and its application to the accelerator structure is given here.

2. The difference equation

Discussing the influence of the transverse wakefield, the TM110 mode is the one that contributes the most. Differences of each model exist in the factors taken into consideration that affect the parameters of the TM110 mode. In the double circuit model, not only the coupling between the adjacent two cells, but also the effects between the TM110 and TE111 modes are also considered. By expanding the fields in each cell into an infinite set of orthonormal modes and relating the coefficients in adjacent cells to one another by treating the iris coupling using Bethe's static approximation, we can get the difference equations for the double circuit [1]:

$$(x_m - \lambda)f_m - \frac{k_{m+1/2}}{2}f_{m+1} - \frac{k_{m-1/2}}{2}f_{m-1} = -\frac{\sqrt{k_{m+1/2}\hat{k}_{m+1/2}}}{2}\hat{f}_{m+1} + \frac{\sqrt{k_{m-1/2}\hat{k}_{m-1/2}}}{2}\hat{f}_{m-1}$$

(2.1)

$$(\hat{x}_m - \lambda)\hat{f}_m + \frac{\hat{k}_{m+1/2}}{2}\hat{f}_{m+1} + \frac{\hat{k}_{m-1/2}}{2}\hat{f}_{m-1} = \frac{\sqrt{k_{m+1/2}\hat{k}_{m+1/2}}}{2}f_{m+1} - \frac{\sqrt{k_{m-1/2}\hat{k}_{m-1/2}}}{2}f_{m-1}$$

(2.2)

where $x_m$, $k_{m\pm1/2}$, $f_m$ are parameters of the TM110 mode, and $\hat{x}_m$, $\hat{k}_{m\pm1/2}$, $\hat{f}_m$ are parameters of the TE110 mode. An equivalent circuit model of Eqs.2.1 and 2.2 is shown in Fig. 1[2], where $x_m = \dfrac{1}{\omega_m^2}, \hat{x}_m = \dfrac{1}{\hat{\omega}_m^2}$ are the resonant frequency of loop m of the TM mode and TE mode respectively.

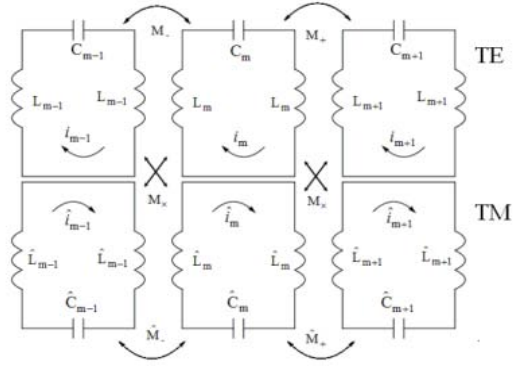

Fig. 1 Equivalent double circuit LC loop of accelerator structure

The double circuit model takes loop m of the chain to represent cell m of the cavity, and the currents in the two chains vary in time as $e^{j\omega t}, e^{j\hat{\omega} t}$, with $\omega, \hat{\omega}$ the angle resonant frequency of the two chains respectively and t the time, and takes the currents in the frequency domain as $i_m(\omega), \hat{i}_m(\hat{\omega})$.

First, the parameters of the eq.2.1 and 2.2 are derived from the dispersion curve of each cell. The dispersion curve for the periodic structure corresponding to the double circuit model is obtained by setting:

$$f_m = f_0 \cos(m\phi) \quad \text{and} \quad \hat{f}_m = \hat{f}_0 \sin(m\phi) \qquad (2.3)$$

In Eqs.2.1 and 2.2, taking all parameters independent of m, Eqs.2.1 and 2.2 become:

$$(x - \lambda - k\cos\phi)f = -\sqrt{k\hat{k}}\sin\phi \ \hat{f} \qquad (2.4)$$

$$(\hat{x} - \lambda + \hat{k}\cos\phi)\hat{f} = -\sqrt{k\hat{k}}\sin\phi \ f \qquad (2.5)$$

Combining the two equations above, the dispersion relation of the two modes is:

$$\cos\phi = \dfrac{k\hat{k} - (x-\lambda)(\hat{x}-\lambda)}{(x-\lambda)\hat{k} - (\hat{x}-\lambda)k} \qquad (2.6)$$

When the phase shift Φ equals zero or π, derived from eq. 2.6, the frequencies of the zero and π modes of the two bands, which can be obtained by the simulation software CST, are given by:

$$\lambda_0^{(1,2)} = x - k, \hat{x} + \hat{k} \quad \text{and} \quad \lambda_\pi^{(1,2)} = x + k, \hat{x} - \hat{k} \qquad (2.7)$$

The four parameters $x, \hat{x}, k, \hat{k}$ can be obtained directly from Eq. 2.7. Given the boundary condition corresponding to a structure with N full cells:

$$f_0 = f_1, f_{N+1} = f_N, \kappa_{1/2} = \kappa_1, \kappa_{N+1/2} = \kappa_N$$
$$\hat{f}_0 = -\hat{f}_1, \hat{f}_{N+1} = -\hat{f}_N, \hat{\kappa}_{1/2} = \hat{\kappa}_1, \hat{\kappa}_{N+1/2} = \hat{\kappa}_N \quad (2.8)$$

The eigen-functions of the double circuit model can be then determined by:

$$\begin{pmatrix} M & M_x \\ M_x^T & \hat{M} \end{pmatrix} \begin{pmatrix} f \\ \hat{f} \end{pmatrix} = \lambda \begin{pmatrix} f \\ \hat{f} \end{pmatrix} \quad (2.9)$$

where:

$$M = \begin{pmatrix} \frac{1}{\omega_1^2} - \frac{\kappa_{1/2}}{2} & -\frac{\kappa_{3/2}}{2} & 0 & 0 & \cdots & 0 \\ -\frac{\kappa_{3/2}}{2} & \frac{1}{\omega_2^2} & -\frac{\kappa_{5/2}}{2} & 0 & \cdots & 0 \\ 0 & -\frac{\kappa_{5/2}}{2} & \frac{1}{\omega_3^2} & -\frac{\kappa_{7/2}}{2} & \cdots & 0 \\ \vdots & \vdots & \vdots & \vdots & \ddots & \vdots \\ 0 & 0 & 0 & \cdots & -\frac{\kappa_{N-1/2}}{2} & \frac{1}{\omega_N^2} - \frac{\kappa_{N+1/2}}{2} \end{pmatrix} \quad (2.10)$$

$$\hat{M} = \begin{pmatrix} \frac{1}{\hat{\omega}_1^2} + \frac{\hat{\kappa}_{1/2}}{2} & \frac{\hat{\kappa}_{3/2}}{2} & 0 & 0 & \cdots & 0 \\ \frac{\hat{\kappa}_{3/2}}{2} & \frac{1}{\hat{\omega}_2^2} & \frac{\hat{\kappa}_{5/2}}{2} & 0 & \cdots & 0 \\ 0 & \frac{\hat{\kappa}_{5/2}}{2} & \frac{1}{\hat{\omega}_3^2} & \frac{\hat{\kappa}_{7/2}}{2} & \cdots & 0 \\ \vdots & \vdots & \vdots & \vdots & \ddots & \vdots \\ 0 & 0 & 0 & \cdots & \frac{\hat{\kappa}_{N-1/2}}{2} & \frac{1}{\hat{\omega}_N^2} + \frac{\hat{\kappa}_{N+1/2}}{2} \end{pmatrix} \quad (2.11)$$

$$M_x = \begin{pmatrix} -\frac{\sqrt{k_{1/2}\hat{k}_{1/2}}}{2} & \frac{\sqrt{k_{3/2}\hat{k}_{3/2}}}{2} & 0 & 0 & \cdots & 0 \\ -\frac{\sqrt{k_{3/2}\hat{k}_{3/2}}}{2} & 0 & \frac{\sqrt{k_{5/2}\hat{k}_{3/2}}}{2} & 0 & \cdots & 0 \\ 0 & -\frac{\sqrt{k_{5/2}\hat{k}_{5/2}}}{2} & 0 & \frac{\sqrt{k_{7/2}\hat{k}_{7/2}}}{2} & \cdots & 0 \\ \vdots & \vdots & \vdots & \vdots & \ddots & \vdots \\ 0 & 0 & 0 & 0 & -\frac{\sqrt{k_{N-1/2}\hat{k}_{N-1/2}}}{2} & \frac{\sqrt{k_{N+1/2}\hat{k}_{N+1/2}}}{2} \end{pmatrix}$$

(2.12)

Written in a more compact manner, Eq.2.9 simply becomes:

$$\tilde{M}\tilde{f} = \lambda \tilde{f} \quad (2.13)$$

By Eq. 2.13, the eigenmodes of the accelerator structure equivalent to the double circuit model can be found.

3. The kick factors

Assuming that the cavity is empty at t=0, and the exciting charge reaches the center of cell m at t=mL/c, then the charge in the cell m at time t is:

$$Q(t) = q\delta(t - mL/c) \quad (3.1)$$

Then in the frequency domain, by the Fourier transform, Eq3.1 becomes:

$$Q(\omega) = q\exp(-jm\omega L/c) \quad (3.2)$$

Then the driving terms in the structure has the effect of adding

$$h_m = \frac{q}{2j\omega C_m \omega_m \sqrt{L_m}} \exp(-jm\omega L/c) = \frac{q}{\omega\sqrt{2C_m}} \exp(-jm\omega L/c - j\pi/2) \quad (3.3)$$

$$\hat{h}_m = \frac{q}{2j\omega \hat{C}_m \hat{\omega}_m \sqrt{\hat{L}_m}} \exp(-jm\omega L/c) = \frac{q}{\omega\sqrt{2\hat{C}_m}} \exp(-jm\omega L/c - j\pi/2) \quad (3.4)$$

to the right of Eqs.2.1 and 2.2 respectively.

For the p$^{th}$ eigenvalue, Eq. 2.13 is written as

$$\tilde{M}\tilde{f}^p = \lambda_p \tilde{f}^p \quad (3.5)$$

where $\tilde{M}$ is the matrix of the system, $\tilde{f}^p$ the p$^{th}$ eigenfunction and $\lambda_p = \omega_p^{-2}$ the p$^{th}$ eigenvalue.

When the exciting charge enters this system, then the eigenfunction becomes:

$$M\tilde{g} - \lambda\tilde{g} = \tilde{h} \quad (3.6)$$

Combining with Eq. 3.5, this leads to:

$$\tilde{g} = \sum_p \frac{\tilde{f}^{(p)}}{\lambda_p - \lambda}\left[\frac{\tilde{f}^{(p)} \bullet \tilde{h}}{|\tilde{f}^{(p)}|^2}\right] \quad (3.7)$$

Going back to the time domain,

$$G_m(t) = \int_{-\infty}^{\infty} d\omega g_m(\omega)e^{j\omega t} = \int_{-\infty}^{\infty} d\omega \sum_p \frac{\tilde{f}_m^{(p)}}{\lambda_p - \lambda}\left[\frac{\sum_n f_n^{(p)} h_n + \hat{f}_n^{(p)}\hat{h}_n}{|\tilde{f}^{(p)}|^2}\right]e^{j\omega t} \quad (3.8)$$

Then:

$$G_m(t) = q\sum_p \frac{\tilde{f}_m^{(p)}\omega_p^2}{|\tilde{f}^{(p)}|^2}\sum_n (\frac{f_n^{(p)}}{\sqrt{2C_n}} + \frac{\hat{f}_n^{(p)}}{\sqrt{2\hat{C}_n}})\int_{-\infty}^{\infty}\frac{\omega d\omega}{\omega^2 - \omega_p^2}e^{j\omega(t-nL/c)-j\pi/2} \quad (3.8)$$

The integral in the above equation contains two poles, at $\omega = \pm\omega_p$. According to the residual theorem [4], the result is:

$$G_m(t) = -2\pi q \sum_p \frac{f_m^{(p)} \omega_p}{|\tilde{f}^{(p)}|^2} \sum_n (\frac{f_n^{(p)}}{\sqrt{2C_n}} + \frac{\hat{f}_n^{(p)}}{\sqrt{2\hat{C}_n}}) \sin\{\omega_p(t-nL/c)\} \quad (3.9)$$

In addition, the current in the $m^{th}$ circuit corresponding to the TM110 mode is [1]:

$$I_m(t) = \sqrt{2C_m} G_m(t) \quad (3.10)$$

The voltage drop caused by the exciting charge in the loop m is:

$$V_{zm}(t) = \frac{1}{C_m} \int_0^t I_m(t') dt' + \frac{q}{C_m} \quad (3.11)$$

Combining Eqs.3.09~3.11, the voltage drop across the capacitor is:

$$V_{zm}(t) = \frac{q}{\sqrt{C_m}} \sum_p \frac{f_m^{(p)}}{|\tilde{f}^{(p)}|^2} \sum_n (\frac{f_n^{(p)}}{\sqrt{C_n}} + \frac{\hat{f}_n^{(p)}}{\sqrt{\hat{C}_n}}) \cos(\omega_p(t-nL/c)) \quad (3.12)$$

The voltage drop across the capacitor represents the energy loss of a test particle to the dipole modes of the cavity. By using the Panofsky-Wenzel theorem [3], the transverse kick can be obtained from the longitudinal one,

$$V_m(t) = (c/x) \int_0^t V_{zm}(t') dt' \quad (3.13)$$

Thus, the transverse kick in cell m is:

$$V_m(t) = \frac{q}{\sqrt{C_m}} \sum_p \frac{f_m^{(p)}}{\omega_p |\tilde{f}^{(p)}|^2} \sum_n (\frac{f_n^{(p)}}{\sqrt{C_n}} + \frac{\hat{f}_n^{(p)}}{\sqrt{\hat{C}_n}}) \sin(\omega_p(t-nL/c)) \quad (3.14)$$

Besides, the value of the capacitor is assumed to be

$C_m = (2K_{s,//}^m x_e xL)^{-1}$, where $K_{s,//}^m = \omega_s^{(m)} K_s^{(m)}/c$ (3.15) is the loss factor of the dipole mode.

So Eq. 3.14 becomes:

$$V_m(t) = 2q_e x_e L \sum_p \frac{f_m^{(p)} \sqrt{K_s^{(m)} \omega_s^{(m)}}}{\omega_p |\tilde{f}^{(p)}|^2} \sum_n (f_n^{(p)} \sqrt{K_s^{(n)} \omega_s^{(n)}} + \hat{f}_n^{(p)} \sqrt{\hat{K}_s^{(n)} \hat{\omega}_s^{(n)}}) \sin(\omega_p(t-nL/c))$$

(3.15)

By the trigonometric function, Eq. 3.14 can be written in terms of amplitude and phases of the modes:

$$V_m(t) = \sum_p V_m^{(p)} \sin(\omega_p t - \theta_p) \quad (3.16)$$

where

$$V_m^{(p)} = \frac{2qxLf_m^{(p)} \sqrt{K_s^{(m)} \omega_s^{(m)}}}{N\omega_p |\tilde{f}^{(p)}|^2} |\sum_n (f_n^{(p)} \sqrt{K_s^{(n)} \omega_s^{(n)}} + \hat{f}_n^{(p)} \sqrt{\hat{K}_s^{(n)} \hat{\omega}_s^{(n)}}) e^{in\varphi_p}|$$

(3.17), and

$$\theta_p = \frac{\sum_n (f_n^{(p)} \sqrt{K_s^{(n)} \omega_s^{(n)}} + \hat{f}_n^{(p)} \sqrt{\hat{K}_s^{(n)} \hat{\omega}_s^{(n)}}) \sin(n\varphi_p)}{\sum_n (f_n^{(p)} \sqrt{K_s^{(n)} \omega_s^{(n)}} + \hat{f}_n^{(p)} \sqrt{\hat{K}_s^{(n)} \hat{\omega}_s^{(n)}}) \cos(n\varphi_p)}$$
(3.18)

In space coordinates with a test particle following a distance s behind the driving charge, the transverse kick is given by:

$$V_m(s) = V_m(\frac{mL+s}{c})$$
(3.19)

Summing up the contribution of all the cavity cells, the total kick felt by the test charge is:

$$V(s) = \sum_m V_m(s)$$
(3.20)

Combining Eq. 3.18~3.20 the total kick felt by the test charge going through the whole structure is:

$$V(s) = \sum_m \sum_p V_m^{(p)} \sin(\omega_p s/c + mL\omega_p/c - \theta_p)$$
(3.21)

By the same operation used to derive Eq.3.15, Eq. 3.21 becomes

$$V(s) = \sum_p \sin(\omega_p s/c + \theta'_p - \theta_p) | \sum_m V_m^{(p)} e^{im\varphi_p} |$$
(3.22)

Where

$$\theta'_p = \frac{\sum_m f_m^{(p)} \sqrt{K_s^{(m)} \omega_s^{(m)}} \sin(m\varphi_p)}{\sum_m f_m^{(p)} \sqrt{K_s^{(m)} \omega_s^{(m)}} \cos(m\varphi_p)}$$
(3.23)

Assuming that the couplings are small so that the effects of any precursor voltage can be ignored, and then when s<0, $\tilde{V}(s) = 0$, which then leads to

$$\tilde{V}(s) \approx 2q_e x_e NL \sum_p K_p \sin\frac{\omega_p s}{c}$$
(3.24)

Where $\varphi_p = \omega_p L/c$, and

$$K_p = \frac{|\sum_n f_n^{(p)} \sqrt{K_s^{(n)} \omega_s^{(n)}} e^{in\varphi_p}| * |\sum_n (f_n^{(p)} \sqrt{K_s^{(n)} \omega_s^{(n)}} + \hat{f}_n^{(p)} \sqrt{\hat{K}_s^{(n)} \hat{\omega}_s^{(n)}}) e^{in\varphi_p}|}{N\omega_p |\tilde{f}^{(p)}|^2}$$
(3.25)

The wakefield function of the entire structure is then given by:

$$W(s) \approx 2\sum_p K_p \sin\frac{\omega_p s}{c}$$
(3.26)

When taking the effects of the TE111 mode into consideration, the kick factor

should be

$$K_p = \frac{|\sum_n (f_n^{(p)}\sqrt{K_s^{(n)}\omega_s^{(n)}} + \hat{f}_n^{(p)}\sqrt{\hat{K}_s^{(n)}\hat{\omega}_s^{(n)}})e^{in\varphi_p}|^2}{N\omega_p |\tilde{f}^{(p)}|^2} \quad (3.27)$$

Considering the definition of shunt impedance [5] and Eq. 3.15

$$K_p = \frac{|\sum_n f_n^{(p)}\sqrt{K_s^{(n)}\omega_s^{(n)}}(1+\varepsilon_n^{(p)})e^{in\varphi_p}|^2}{N\omega_p |\tilde{f}^{(p)}|^2} \quad (3.28)$$

Where $\varepsilon_n^{(p)} = \dfrac{\hat{f}_n^{(p)}}{f_n^{(p)}} \dfrac{\hat{V}_{s,//}^{(n)}}{V_{s,//}^{(n)}}$ (3.29)

4. Wakefield calculation results

Using Eq.3.25-3.26, the distribution of the wakefield parameters and its envelope are calculated, giving the results shown in Figures 2 and 3.

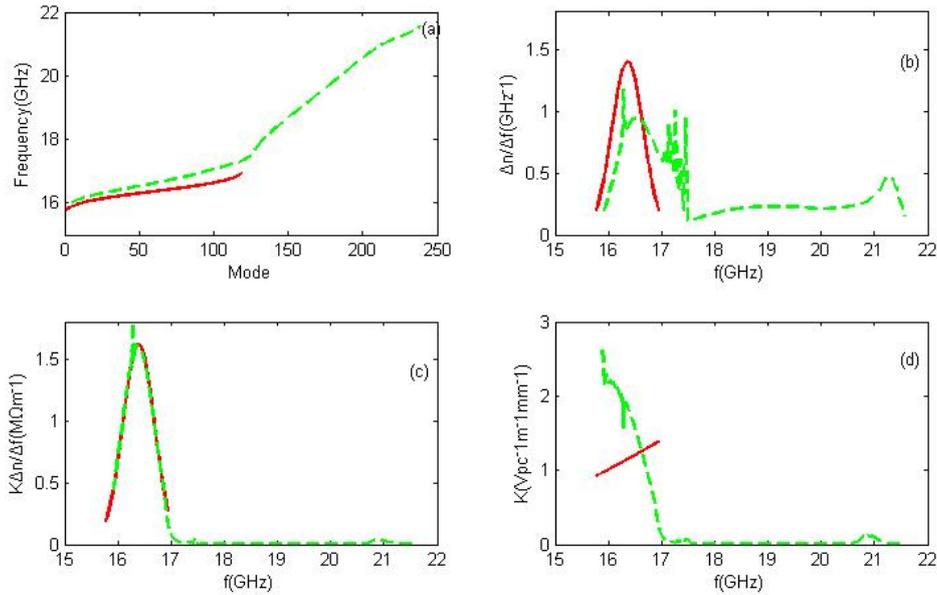

Figure 2: Results of the double circuit model for: (a) the mode spectrum, (b) the mode density (normalized), (c) the product of kick factor and mode density, (d) the kick factor. Red lines show the results of the uncoupled model and green dashed lines show the results of the double circuit model for comparison.

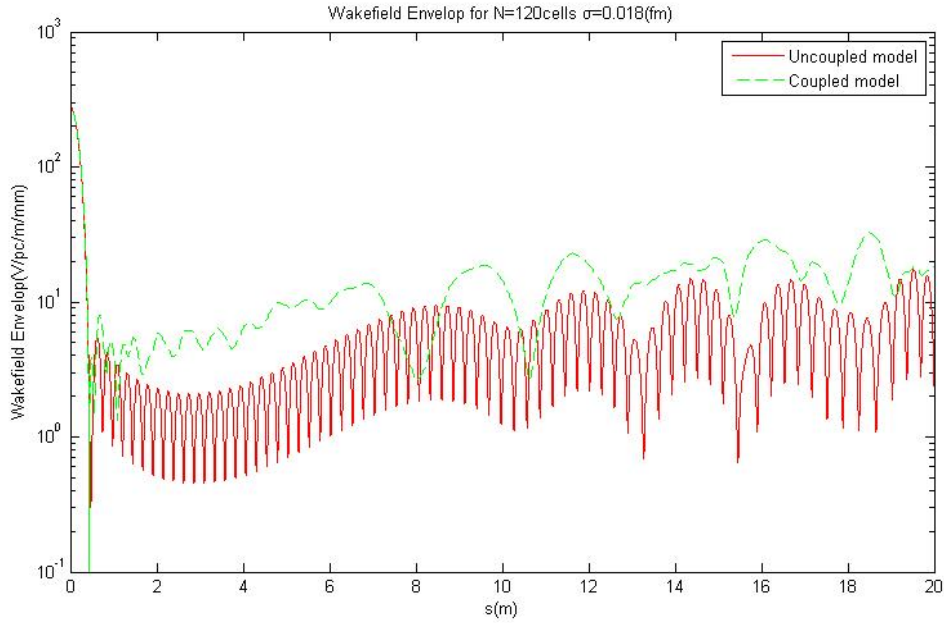

Figure3. The envelope of the wake function calculated using the uncoupled model (red line) and the double circuit model (green dashed line).

From Fig. 2, modes of the second band (roughly over 17 GHz) have low kick factors and therefore do not contribute significantly to the wakefield. As can be seen in Fig3, there are some differences between the results of the uncoupled model and the double circuit model, but they have the same general tendency.

5. Conclusion

On the basis of the eigen-functions of the equivalent double circuit model, expressions to calculate the long range transverse wakefield and its parameters, the mode frequency and kick factor, have been deduced, as shown in Eqs.3.24-3.29. The results of those parameters of the accelerator structure designed, calculated by the deduced expressions, have been given in the end, with a comparison with the results of the uncoupled model. From the comparison of the results of the two different models, the double equivalent circuit model do give some more informations about the TE111 mode, which gives secondary impact on the following bunches, just as the result shows. On the other hand, the comparison of the wakefield envelope not only gives a support of the second model, but also shows the effects of the TE111 mode to the distribution of the wakefield envelope.